\documentclass[12pt,prb,reprint]{revtex4-1}
\usepackage{xcolor}
\usepackage{amsmath}
\usepackage{graphicx}
\usepackage[normalem]{ulem}

\begin{document}

\title{Generating derivative superstructures for systems with high configurational freedom}

\author
{Wiley S. Morgan}
\affiliation
{Department of Physics and Astronomy, Brigham Young University, Provo Utah 84602 USA}

\author
{Gus L. W. Hart}
\affiliation
{Department of Physics and Astronomy Brigham Young University, Provo Utah 84602 USA}

\author 
{Rodney W. Forcade}
\affiliation
{Department of Mathematics Brigham Young University, Provo Utah 84602 USA}

\begin{abstract}


  Modeling potential alloys requires the exploration of all possible
  configurations of atoms. Additionally, modeling the thermal
  properties of materials requires knowledge of the possible ways of
  displacing the atoms. One solution to finding all symmetrically
  unique configurations and displacements is to generate the complete
  list of possible configurations and remove those that are
  symmetrically equivalent. This approach, however, suffers from the
  combinatorial explosion that happens when the supercell size is
  large, when there are more than two atom types, or when there are
  multiple displaced atoms. This problem persists even when there are
  only a relatively small number of unique arrangements that survive
  the elimination process. Here, we extend an existing
  algorithm\cite{enum1,enum2,enum3} to include the extra
  configurational degree of freedom from the inclusion of displacement
  directions. The algorithm uses group theory to eliminate large
  classes of configurations, avoiding the combinatoric explosion. With
  this approach we can now enumerate previously inaccessible systems,
  including atomic displacements.

\end{abstract}

\maketitle

\section{Introduction} \label{Intro}

In computational material science, one frequently needs to list the
``derivative superstructures''\cite{Buerger:1947fr} of a given
lattice.  A derivative superstructure is a structure with lattice
vectors that are multiples of a ``parent lattice'' and have atomic
basis vectors constructed from the the lattice points of the parent
lattice. For example, many phases in metal alloys are merely
``superstructures'' of fcc, bcc, or hcp lattices (L1$_{0}$, L1$_{2}$,
B2, D0$_{19}$, etc.). When modeling alloys it is necessary to explore
all possible configurations and concentrations of atoms within these
superstructures. When determining if a material is thermodynamically
stable, the energies of the unique arrangements are compared to
determine which has the lowest energy.

Derivative superstructures are found using combinatoric
searches\cite{alg5,alg1,alg2,alg4,enum3,enum2,enum1}, comparing every
possible combination of atoms to determine which are unique. However,
these searches can be computationally expensive for systems with high
configurational freedom and are sometimes impractical due to the large
number of possible arrangements.

The inefficiency of combinatoric searches makes finding the unique
derivative superstructures a limiting factor in searches for high
entropy alloys (HEA)\cite{HEA1,HEA2,HEA3}. The configurational
complexity of HEAs prevents them from phase separating; this same
complexity makes listing every possible arrangement of atoms
impractical with current algorithms.

Other problems impaired by the inefficiency of current enumeration
methods include modeling materials that have disorder in their
structures, such as site-disordered solids\cite{site_disorder} or
that include atomic displacements as a degree of freedom, such as
phonon models.\cite{Vidvuds:2014jh,Doak:2015jd} There are numerous
techniques available for modeling these systems including cluster
expansion (CE)\cite{Sanchez:1984bh} and a recently developed ``small
set of ordered structures'' (SSOS) method.\cite{Jiang:2016fo} However,
the accuracy of these methods is still linked to the number of unique
configurations being modeled. In other words, if the model is trained
on a small set of configurations then it will not be able to make
accurate predictions. Increasing the number of configurations used to
train the models can improve their predictive powers. Increasing the
number of structures being used requires a more efficient enumeration
technique than those currently available.

Leveraging the basic concepts of the algorithm presented in
Ref. \citenum{enum3}, we altered the algorithm to have more
favorable scaling in multinary cases.  The basic idea is to imagine
the enumeration as a tree search and employ two new ideas: (1)
``partial colorings'' and (2) stabilizer subgroups. Sec.~\ref{Tree}
illustrates the algorithm with a concrete example.

\begin{figure*}
  \includegraphics[scale=0.5]{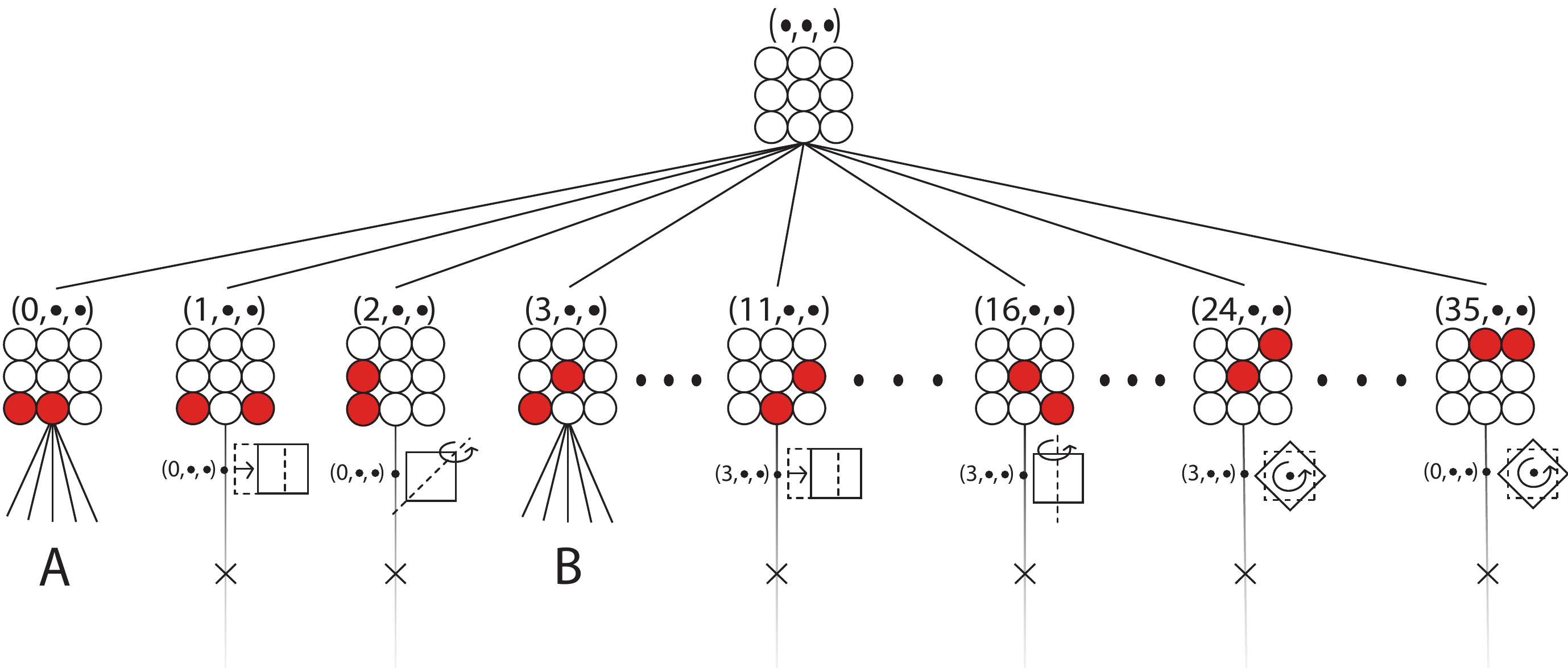}
  \caption{(Color online) The empty lattice and 8 of the 36
    configurations with only red atoms are shown for the example
    discussed in section \ref{Tree}. Above each partial coloring is a
    vector that indicates it's location in the tree,
    i.e. $(x_r,x_y,x_p)$, where the $x_i$s are integers that indicate
    which arrangement of that color is on the lattice and a $\bullet$
    means that no atoms of that color have been placed yet. Below each
    configuration is either the label of a symmetrically equivalent
    configuration, along with the group operation that makes them
    equivalent, or the letters A and B. A and B are the branches that
    are built from the 1-partial colorings that are unique and are
    displayed in Fig.  \ref{fig:master_tree_2}}
  \label{fig:master_tree_1}
\end{figure*}

The concept of partial colorings is to skip entire branches of the
tree that are symmetrically equivalent to previously visited
branches. A partial coloring is an intermediate level in the tree (see
Fig.~\ref{fig:master_tree_1}) where configurations are not yet
completely specified. It frequently happens that symmetric redundancy
can be identified at an early, ``partially colored'' stage, avoiding
the need to descend further down the tree.


Stabilizer subgroups further increase the efficiency of the new
algorithm.  Any symmetrically-equivalent full colorings further down
the current branch will have the same partial coloring. Thus, the only
symmetries that are relevant are those that leave the current partial
coloring unchanged. These symmetries form a (stabilizing) subgroup of
the full group. This significantly impacts the efficiency because the
stabilizer subgroup is often much smaller than the full group.

\section{Supercell selection and the Symmetry Group} \label{supercell}

The first step in enumerating derivative superstructures is the
enumeration of unique supercells. This step has already been solved by
Hart and Forcade\cite{enum1}, but due to its importance to the
algorithm we provide a brief overview.

The supercells, of size n, are found by constructing all Hermite
Normal Form (HNF) matrices whose determinant is $n$. An HNF matrix is
an integer matrix with the following form and relations:
\begin{equation}
  \begin{pmatrix}
    a & 0 & 0 \\
    b & c & 0 \\
    d & e & f
  \end{pmatrix}, 0 \leq b < c, 0 \leq d < f, e < f
\end{equation}
where $acf=n$. The HNFs determine all possible the supercells for the
system. For example, consider a 9-atom cell, then $n=9$ and $a$,
$c$, $f$ are limited to permutations of (1,3,3) and (1,1,9). Then
following the rules for the values of $b$, $d$, and $e$, every HNF for
this system can be constructed. These HNFs represent all the possible
supercells of size $n$ of the selected lattice. Some of these are
equivalent by symmetry, so the symmetry group of the parent lattice is
used to eliminate any duplicates.

Next, we convert the symmetries of the lattice to a list of
permutations of atomic sites. There is a one-to-one mapping between
the symmetries of the lattice and atomic site permutations, i.e., the
groups are isomorphic. The mapping from the symmetry operations to the
permutation group is accomplished using the quotient group $G = L/L'$,
where $L$ is the lattice, constructed from the unit cell, and $L'$ is
the superlattice, constructed from the supercell. The quotient group
$G$ is found directly from the Smith Normal Form (SNF) matrices, which
can be constructed from the HNFs via a standard algorithm using
integer row and column operations. Thus $S=UHV$ where $U$ and $V$ are
integer matrices with determinant $\pm 1$ and $S$ is the diagonal SNF
matrix, where each positive integer diagonal entry divides the next
one down. The group, $G$, is then $G=Z_{s_1} \bigoplus Z_{s_2}
\bigoplus Z_{s_3}$, where $s_i$ is $i$th diagonal of the SNF and
$Z_{s_i}$ represents the cyclic group of order $n$.

Once the supercells have been found and their symmetry groups have
been converted to the isomorphic permutation group, the algorithm can
begin finding the unique arrangements of atoms within each supercell
in a tree search framework. This is accomplished by treating each
supercell with its symmetry group as a separate enumeration
problem. The results of the enumeration across all supercells are
then combined to produce the full enumeration.

\section{Tree Search} \label{Tree}

Once a supercell has been selected, the remainder of the enumeration
algorithm resembles a tree search in which each branch corresponds to
a specific configuration of atoms within the supercell, many of which
are not fully populated and are called partial colorings (see
Fig. \ref{fig:master_tree_2}). The partial colorings are identified
using a vector that indicates their locations within the tree. Once a
partial coloring is constructed, the stabilizer subgroup for that
partial coloring is found. The stabilizer subgroup allows for the
comparison of branches within the tree in a manner that minimizes the
number of group operations used. These tools, (partial colorings and
the stabilizer subgroup), are used to ``prune'' branches of the tree
as they are being constructed, eliminating large classes of
arrangements at once.

We will use a 2D lattice of 9 atoms as an illustrative example of the
algorithm. The lattice will be populated with the following atomic
species; 2 red atoms, 3 yellow atoms, and 4 purple atoms. A subset of
the possible arrangements of this system is shown in
Fig. \ref{fig:master_tree_2}. The concepts illustrated with this 2D
example are equally applicable in 3D.

\subsection{Partial Colorings} \label{Partials}

\begin{figure*}
  \includegraphics[scale=0.65]{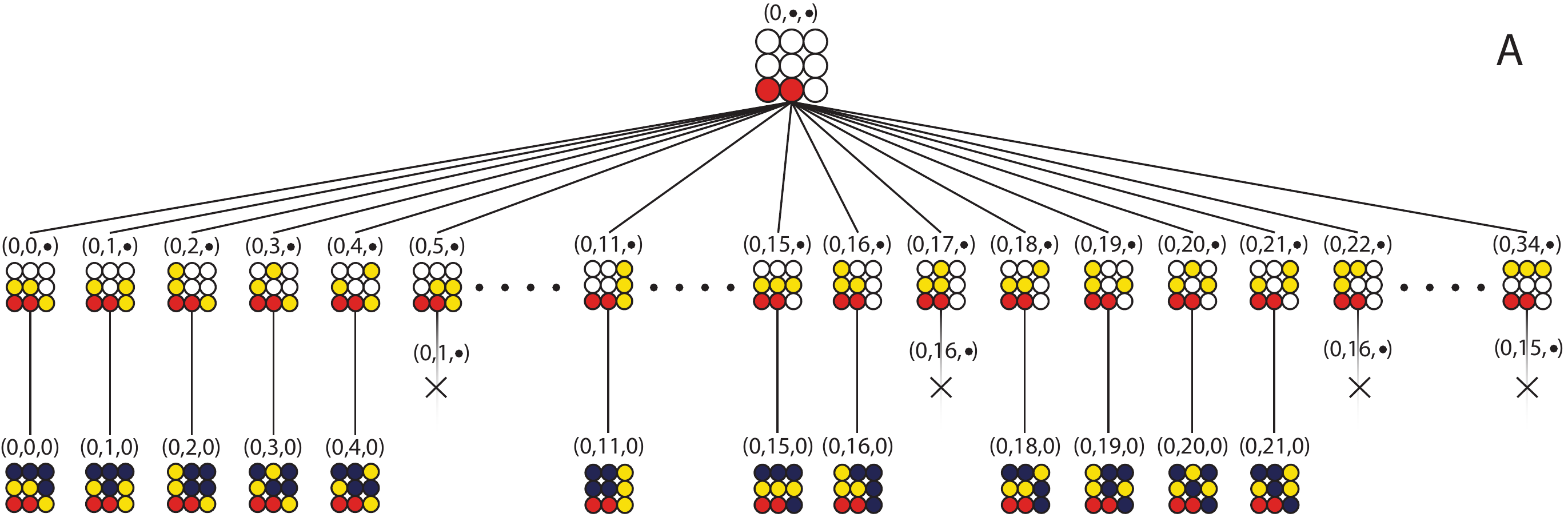}
  \newline
  \newline
  \newline
  \includegraphics[scale=0.65]{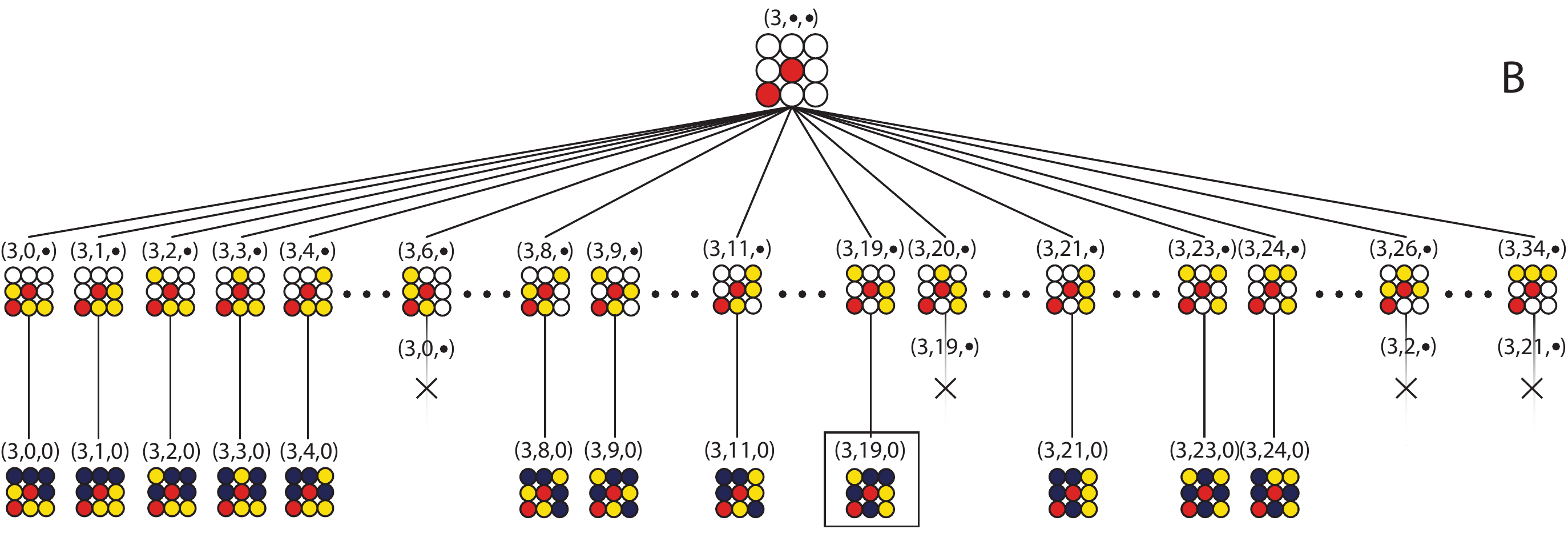}
  \caption{(Color online) Here the A and B branches of the tree from
    Fig. \ref{fig:master_tree_1} are shown. Each branch starts with
    the initial 1-partial coloring the branch is built from
    ($(0,\bullet,\bullet)$ and $(3,\bullet,\bullet)$
    respectively). The branches then show a selection of the 2-partial
    colorings for that branch, and the unique full colorings that are
    found. As in Fig. \ref{fig:master_tree_1} the vectors that
    indicate the configurations location in the tree are displayed
    above the configurations and the symmetrically equivalent labels
    appears beneath them. In this figure the actions that make the
    configurations have been excluded due to their complexity. For
    example, The configuration labeled $(0,5,\bullet)$ is equivalent
    to the $(0,1,\bullet)$ configuration by a rotation about the
    vertical followed by a translation to the left. In the B branch
    configuration $(3,19,0)$ is outlined for reference because it is
    used as an example later in the text.}
  \label{fig:master_tree_2}  
\end{figure*}

When searching for all unique configurations, it is useful to know, a
priori, how many configurations are expected. A recently developed
numerical algorithm for the P\'olya enumeration
theorem\cite{Polya:1987,Polya:1937,polya} allows one to quickly
determine the memory requirements of storing the unique
arrangements. For the 9 atom system considered here, the P\'olya
algorithm predicts that there are 24 unique arrangements to be found.




The algorithm places atomic species on the lattice according to their
concentrations. In this case, the red atoms have the lowest
concentration and are placed in the first two sites of the cell
creating the first 1-partial coloring (a partial coloring is a
configuration with only a subset of the atoms decorating the
lattice). This is shown in the leftmost configuration, labeled
$(0,\bullet,\bullet)$, in the second row of
Fig. \ref{fig:master_tree_1}. The general procedure is to apply the
symmetry group to each partial coloring in order to make comparisons
between partial colorings and determine if they are symmetrically
equivalent. For example, in Fig. \ref{fig:master_tree_1}, the
configuration labeled $(1,\bullet,\bullet)$ is equivalent to
configuration $(0,\bullet,\bullet)$ by a translation of the
lattice. At this stage we only have one partial coloring so it is
unique and no comparisons need to be made, however the symmetry group
is still applied to find the stabilizer subgroup described in section
\ref{Stabs}.


Comparisons between configurations are made by using a hash
function. In computer science, any data set can be placed in a hash
table which associates a hash, or label, with the data. In our case,
the configurations are listed within the hash table in the order they
are created. The hash function then maps the configuration to a vector
of integers with an entry for each species, color, in the system. The
hash function used is similar to the one described in
Ref. \citenum{enum3}. However, due to its importance in this
algorithm, we provide an overview of how the hash function works.

The hash function for the algorithm uses the principles of
combinatorics to uniquely identify each partial coloring using an
integer vector. Its construction starts by determining the number of
possible ways to arrange the colors on the lattice. The number of
possible configurations can be found using the multinomial
coefficient, which is equivalent to the product of binomial
coefficients for each individual color:
\begin{equation} 
  \begin{split}
  C = {n \choose a_1,a_2,..,a_k} = C_1C_2...C_k =\\
  {n \choose a_1}{n-a_1 \choose a_2}...{n-a_1-a_2-...-a_k \choose a_k},
  \end{split}
  \label{eq:multinomial}  
\end{equation}
where $n$ is the number of sites in the unit cell and
$a_1,a_2,...,a_k$ are the number of atoms of species $i$ such that
$\sum_{i}a_i = n$.  The binomials determine the number of ways to
place the atoms of each color within the lattice once the previous
colors have been placed. By assigning each partial coloring an
integer, $x_i$, from 0 to $C_i-1$, where $i$ is the color, we can
build a vector that identifies the location, $(x_1,x_2,...,x_k)$, of
the configuration within the tree. For example, there are $C_r = {9
  \choose 2} = 36$ ways to place the red atoms on the empty
lattice. After the red atoms are placed then there remain $C_y = {7
  \choose 3} = 35$ ways to place the yellow atoms on the remaining
lattice sites. This leaves $C_p = {4 \choose 4} = 1$ way to place the
purple atoms on the lattice. Within Fig. \ref{fig:master_tree_1} and
\ref{fig:master_tree_2}, the vector locations have the form
$(x_r,x_y,x_p)$ and if the color has not been assigned yet then the
$x_i$s are replaced by dots indicating an empty vector site.

The hash function is a one-to-one mapping between the configurations
to the location vectors. These numbers are constructed by considering
each color separately and building a binary string of the color and
the remaining empty lattice sites, where the color is a 1 and the
empty site is a 0 within the string. From the binary string, we can
then use a series of binomial coefficients to find the $x_i$'s. The
binomial coefficients are found by taking each 0 in the string that
has 1's to the right of it and computing ${p \choose q-1}$, where $p$
is the number of digits to the right and $q$ is the number of 1's to
the right of the 0. Summing the binomials for qualifying zero produces
a number that tells us how many configurations came before the current
one.

As an example of the hash function that constructs the location
vector, consider configuration (3,19,0) of
Fig. \ref{fig:master_tree_2}B. The construction begins with the red
atoms represented as the following binary string (1,0,0,0,1,0,0,0,0),
where every atom that is not red has been represented by a 0 and the
red atoms by a 1. This string has 3 zeros that have a single 1 to
their right, the first zero has 7 digits to its right, the second has
6 atoms to its right and the third has 5 atoms to its right. The
resultant sum of binomials is $x_r={7 \choose 0} + {6 \choose 0} + {5
  \choose 0} = 1 + 1 + 1 = 3$. This result is the first entry in our
location vector.

The second entry in the location vector is constructed for the yellow
atoms. The bit string representation of the yellow atoms is
(0,1,0,1,1,0,0), there are only 7 digits because the 2 red atoms have
already been placed, so $x_y={6 \choose 2} + {4 \choose 1} = 15 + 4 =
19$. The last entry in the location vector is built for the purple
atoms which have the bit string (1,1,1,1), so $x_p=0$. The location
vector is complete once all atoms within a configuration have been
included.





The location vectors allow us to determine if a configuration is
unique by checking if an element of the symmetry group maps the
configuration to a configuration with a smaller location vector. The
symmetry operations map a configuration's location to a second,
equivalent location. Uniqueness is determined by comparing the
original and mapped locations for the configuration; if the mapped
configuration has already been enumerated, that is, if
$x\mathrm{_{original}} > x\mathrm{_{mapped}}$, then the configuration
is not unique because it is equivalent to one we have already
visited. For example, configuration $(2,\bullet,\bullet)$ shown in
Fig. \ref{fig:master_tree_1} can be turned into configuration
$(0,\bullet,\bullet)$ by a 180 degree rotation about the
diagonal. Since $(2,\bullet,\bullet)$ and $(0,\bullet,\bullet)$ are
equivalent we conclude that $(2,\bullet,\bullet)$ is not unique
because $2 > 0$. In summary, if any element of the symmetry group
makes the location vector ``smaller'', then the corresponding
configuration has already been visited.

\subsection{The Stabilizer Subgroup} \label{Stabs}

The algorithm is efficient because the entire symmetry group does not
need to be applied to a partial coloring, only the stabilizer subgroup
of the partial coloring one level up the tree is needed. The
stabilizer subgroup is found when the symmetry group was applied to
the partial coloring one level higher up the tree, so finding the
stabilizer subgroup costs nothing computationally. As an example of an
element of the stabilizer subgroup, consider the cell
$(3,\bullet,\bullet)$, displayed in Fig. \ref{fig:master_tree_1}, and
reflect it about the diagonal; the red atoms are unaffected. This
means that a reflection about the diagonal is a member of the
stabilizer subgroup for the 1-partial coloring
$(3,\bullet,\bullet)$. In general, only a small subset of the symmetry
group will be in the stabilizer subgroup for any partial coloring.

\begin{figure} 
  \centering
  \includegraphics[scale=0.7]{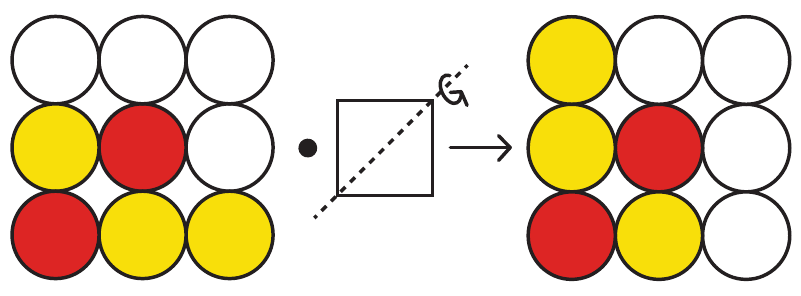}
  \caption{(Color online) The configuration $(3,0,\bullet)$, shown on
    the left, is acted on by a reflection about the diagonal resulting
    in configuration $(3,6,\bullet)$, shown on the right. Because the
    symmetry group operation is a stabilizer for the configuration
    $(3,\bullet,\bullet)$ the red atoms were not affected. A
    stabilizer is a group element that leaves the set invariant. The
    yellow atoms, however, were mapped to a different configuration. This
    means we can use just the stabilizer subgroup for the
    $(3,\bullet,\bullet)$ configuration to compare all the 2-partial
    colorings of the form $(3,x_y,\bullet)$, where ($0 \le x_y \le
    C_y-1$), because any other group operation would map us to a
    different branch of the tree.}
  \label{fig:stabilizer}
\end{figure}

The stabilizer subgroup leaves the desired $n$-partial coloring
unchanged, where $n$ is the depth in the tree. Once another color is
added (making an $(n+1)$-partial coloring) the stabilizer subgroup for
the $n$-partial coloring becomes the only group operations that can be
applied without affecting the $n$-partial coloring. In other words, if
we were to use any other group elements we would be comparing
configurations that we already know are equivalent on the $n$-partial
coloring level.


Once a unique $n$-partial coloring and its stabilizer subgroup have
been found, the algorithm proceeds down the branch to the
$(n+1)$-partial colorings, see Fig. \ref{fig:master_tree_2}. To check
the uniqueness of the $(n+1)$-partial colorings, the stabilizer subgroup
from the $n$-partial coloring are used and the stabilizer subgroup for
the $(n+1)$-partial colorings are stored. When a unique configuration
is found on the $(n+1)$ level another color is added, making the
$(n+2)$-partial colorings, and the process starts over again.

The algorithm proceeds down a branch of the tree until a unique full
configuration is found, such as (0,0,0) of
Fig. \ref{fig:master_tree_2}. When the full configuration is found,
the algorithm backs up one level and considers the next partial
coloring. When no partial colorings are available on a level, the
algorithm backs up until it finds a level with untested partial
colorings. In this manner, the entire tree is explored but only
sections with unique configurations are explored in detail.

For an example of the complete algorithm, consider
Figs. \ref{fig:master_tree_1} and \ref{fig:master_tree_2}. The
algorithm starts at $(\bullet,\bullet,\bullet)$ then builds the
1-partial coloring at $(0,\bullet,\bullet)$, which is unique by virtue
of being the first partial coloring considered on this level, and
records its stabilizer subgroup. The yellow atoms are then added to
the configuration to build the 2-partial coloring at $(0,0,\bullet)$,
of Fig. \ref{fig:master_tree_2} A, which is also unique, and records
its stabilizer subgroup. Next, it places the purple atoms to get the
configuration at (0,0,0); this configuration is saved, then the
algorithm backs up to the 2-partial coloring level to consider the
configuration $(0,1,\bullet)$ and find its stabilizer subgroup.

Once this process has been repeated for all 34 partial colorings in
the vector $(0,x_y,\bullet)$ ($0 \le x_y \le 34=C_y$), the algorithm
retreats to the 1-partial coloring level shown in the second row of
Fig. \ref{fig:master_tree_1} and finds that $(1,\bullet,\bullet)$ and
$(2,\bullet,\bullet)$ are equivalent to $(0,\bullet,\bullet)$. It then
begins to build the $(3,\bullet,\bullet)$ branch, of
Fig. \ref{fig:master_tree_2} B, in the same manner as the
$(0,\bullet,\bullet)$ branch.

Since there are only two unique 1-partial colorings for this system
the algorithm is complete once both branches that originate from
these 1-partial colorings have been explored. In the end, 24 unique
configurations are found (shown in Fig. \ref{fig:master_tree_2}A and
\ref{fig:master_tree_2}B), in agreement with the prediction from the
P\`olya enumeration algorithm.

\subsection{Extension to Include Additional Degrees of Freedom} \label{arrows}

Having established the algorithm, we will now address its extension to
include displacement directions. These enumerations are more difficult
because including displacement directions changes the action of the
group. Displacement directions simply indicate the direction that an
atom could be displaced off the lattice. The enumeration of structures
that include displacement directions can be used to build
databases\cite{Jain2013} of possible structures with displacements
included.

\begin{figure} 
  \centering
  \includegraphics[scale=0.7]{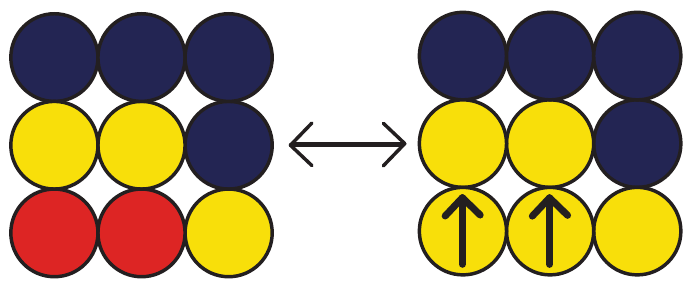}
  \caption{(Color online) To include displacement directions to the
    algorithm we represent the atoms to be displaced by a unique color
    and then convert them back once a unique configuration is
    found. In this figure two displaced yellow atoms are represented
    by red atoms until the previous portion of the algorithm is
    complete, then they are replaced by arrows again for the arrow
    enumeration.}
  \label{fig:arrows&colors}
\end{figure}

Our algorithm changes only slightly if displacement directions are
included in the enumeration. First, the atoms that will be displaced
are treated as a different atomic species so that each displaced
atom's unique locations can be determined (see
Fig. \ref{fig:arrows&colors} for an example where yellow displaced
atoms are replaced with the red atoms from the example system used
above). Once the arrows have been replaced by atomic species, the
algorithm proceeds as normal until a full configuration is found. The
algorithm then restores the arrows and uses the stabilizer subgroup of
the full configuration to check for equivalent arrow configurations.

In order to determine if the combined arrow and color configuration is
unique, each group element has to be paired with a second set of
permutations that determine how the symmetry operation affects the
arrows. The effect on the arrows is represented as a permutation of
the numbers 0 to $d-1$, where each number represents a different
displacement direction up to the $d$ directions being considered. For
example, if we consider the system in Fig. \ref{fig:arrows&colors}, we
have two atoms being displaced along one of the 6 cardinal directions,
then any arrow could have values of between 0 and 5 where each integer
has an associated direction; up=0, right=1, down=2, left=3, into the
page=4, and out of the page=5. The initial arrow vector, shown in the
figure, is (up,up) and is represented as (0,0).


The comparison of the rotated and unrotated arrows is achieved using a
hash function. This function gives each arrow configuration a unique
label that corresponds to the order it is constructed within the
algorithm. This hash function takes a vector of arrow directions
$(a_0,a_1,a_2,....,a_k)$, where $a_i$ is an integer from 0 to $d-1$
indicating the direction of the $i\mathrm{th}$ arrow and $k+1$ is the
number of arrows, and finds:
\begin{equation}
  x_a = \sum_{i=0}^k a_id^i
\end{equation}
This gives each arrow arrangement a unique integer label that we can
then compare between symmetry operations. As was the case for the
configurations, if the effect of a symmetry operation results in a
relationship of $x\mathrm{_{old}} > x\mathrm{_{new}}$, then the the
arrow configuration is not unique and can be ignored.

The stabilizer subgroup for the unique color configuration are used to
map the arrows to new directions and the hash function is used to
compare the original and mapped arrows. After an arrow arrangement is
checked, the algorithm then increases the magnitude of the last $a_k$
in the vector by 1 and checks it for uniqueness with the
stabilizer subgroup. If increasing the magnitude of $a_k$ would cause it to be
greater than the value of $d-1$ then $a_k$ becomes 0 again and
$a_{k-1}$ is increased by 1. This process is repeated until all the
entries in the arrow vector are equal to $d-1$.

For example, the initial arrow vector for the system shown in
Fig. \ref{fig:arrows&colors} is (up,up) and is represented as (0,0).
It is found to be unique since it is the first arrangement. For the
next arrangement the arrow on the right is rotated to point to the
right creating the arrangement represented as (0,1). This arrangement
is also checked to see if it is unique. The right most arrow continues
to be rotated every time a new arrangement is constructed until it is
pointing out of the page and the arrangement represented as (0,5) has
been considered. At this point all possible arrangements that have the
first arrow pointing up have been considered, so the second arrow is
set to point up and first arrow is rotated to make the arrangement
(1,0). We then go back to increasing the last entry in the vector to
create new arrangements in order to determine if any of them are
unique until (1,5) is reached. The process is repeated until all
possible the arrangements, i.e., all 2-tuples of $0...(d-1)$, have
been considered. Once all the vectors have been considered, the
algorithm goes back up the tree to find the next unique configuration
of colors.

In this manner, discrete displacement directions can be added to the
configurations. In this example, adding arrows to the system increases
the number of possible arrangements to 45360 (the number of possible
arrangements for just the atoms is 1260). However, the resultant
number of unique arrangements is only 663.




\section{Conclusion} \label{conclusion}

Our previous algorithms\cite{enum1,enum2,enum3} explored configuration
space by comparing all possible configurations of the atoms to
eliminate those that were symmerically equivalent. These algorithms
were only effective for systems with relatively small amounts
configurational freedom due to the combinatoric explosion that occur
for systems with high configurational freedom and were incapable of
enumerating systems that included displacement directions.

With this new algorithm, it is now possible to find the unique
arrangements of systems with high configurational freedom. The systems
now accessible include $k$-nary alloys and structures with
displacement directions. This is accomplished by using an approach
which closely resembles a tree search, in which large classes of
configurations are eliminated at a time. In this manner, we are able
to avoid the combinatoric explosion which impedes the performance of
the previous algorithms. This algorithm's ability to efficiently
determine the unique arrangements of these systems enables more
effective computational studies.

This research was funded by ONR grant MURI N00014-13-1-0635. This
algorithm has been implemented in the enumlib package and is available
for public use~\footnote{https://github.com/msg-byu/enumlib}.

\section{Acknowledgements}
This work was supported under: ONR (MURI N00014-13-1-0635).

\section{References}

\bibliographystyle{unsrt}

\end{document}